\documentclass[reprint,aps,nofootinbib]{revtex4-2}

\usepackage{amsmath}

\begin{document}
\title{Order-reversed Kubo formulas in relativistic kinetic theory}

\author{Sangyong Jeon}
\email{sangyong.jeon@mcgill.ca}
\affiliation{Department of Physics, McGill University, 3600 rue University, Montreal, H3A2T8 QC, Canada}
\author{Juhee Hong}
\email{juhehong@gmail.com}
\affiliation{Department of Physics and Institute of Physics and Applied Physics, Yonsei University, Seoul 03722, Korea}
\author{Alina Czajka}
\email{alina.czajka@ncbj.gov.pl}
\affiliation{Theoretical Physics Division, National Centre for Nuclear Research, Pasteura 7, Warsaw 02-093, Poland}

\begin{abstract}
Using the stress-energy tensor response functions obtained in the Anderson-Witting model
of kinetic theory with non-zero mass, 
we verify that  these response functions satisfy the required analytic conditions and are
fully consistent with the recently derived Kubo formulas in Ref.\cite{Jeon:2025kem}.

\end{abstract}

\maketitle

\section{Introduction}

In a previous paper \cite{Jeon:2025kem}, we introduced eight new Kubo formulas for 
the shear and bulk viscosities.
Due to the full understanding of the analytic structures of stress-energy correlation
functions in the small $\omega$ and small $k$ limits, we were able to get most of
these Kubo formulas 
by taking the $\omega\to 0$ limit first, instead of taking the usual thermodynamic limit.
Thus, using these formulas 
require stress-energy tensor correlation functions that explicitly exhibit
hydrodynamics poles in the small $\omega$ and $k$ limit.

In the standard transport coefficient calculations in quantum field theory, the 
zero frequency and zero wavenumber limits can be taken without encountering any poles
\cite{Jeon:1994if} because no propagator can diverge in this limit due to the finite thermal width. 
In terms of the kinetic theory language, this procedure corresponds to
first order Chapman-Enskog method. 
Since the limits commute without an intervening pole,
this method cannot be directly used for the new formulas. 
Instead, one should keep $\omega$ and $k$ finite, solve for the stress-energy tensor
correlation functions, and then take the appropriate limits.

Through the studies in
Refs.\cite{Jeon:1994if,Jeon:1995zm,Carrington:1999bw,Wang:1999gv,Wang:2002nba,
Arnold:2000dr,Arnold:2002zm,Arnold:2003zc,Arnold:2006fz},
it has been well established that performing the ladder diagram
resummation
is equivalent to the
linearized kinetic theory. This is true not only in the context of the Chapman-Enskog
method, it can also be shown that when the frequency and the wavenumber are
kept finite, one can recover the linearized kinetic theory equation 
via Kadanoff-Baym, or Schwinger-Dyson equations
\cite{Jakovac:2001kj,Jakovac:2002rc,Arnold:2002ja,Arnold:2002zm,
Fillion-Gourdeau:2006gum,Gagnon:2007qt}.
Therefore, it is worth while to consider the linearized kinetic theory 
as a prelude to the full field theory calculations.

In Refs.\cite{
Romatschke:2015gic,
Hong:2010at,
Hataei:2025mqf,
Bajec:2024jez,
Bajec:2025dqm},
stress-energy tensor correlators have been calculated 
in the linearized kinetic theory within the relaxation time approximation.
Because the full collision integral is not used, one cannot claim that these are fully equivalent
to ladder diagram summations in field theories. Nevertheless, these correlation functions 
do exhibit hydrodynamic poles in the small $\omega$ and $k$ limit. Therefore, it is
relevant to see if these correlation functions satisfy all the conditions expected of the
stress-energy tensor correlation functions, and also to verify the new Kubo formulas.

In this work, we use the relaxation time approximation and work in the high temperature
limit where $m/T \ll 1$.
We leave consideration of the possible space-time dependent quasi-particle mass for future
investigation.

\section{Stress-energy tensor correlators from the kinetic theory}
\label{sec:non_zero_m_kinetic}

For simplicity, we will consider a single component Boltzmann gas in the Anderson-Witting
model \cite{Anderson:1974nyl}.
In this section, we will closely follow the analyses
from Refs.\cite{Romatschke:2015gic,Hataei:2025mqf,Bajec:2025dqm}.
The kinetic theory equation under a small metric perturbation
$h_{\mu\nu} = g_{\mu\nu} - \eta_{\mu\nu}$ is 
\begin{equation}
p^\mu\partial_\mu f - \Gamma^\alpha_{\beta\gamma}{p^\beta p^\gamma\over p^0}
{\partial\over \partial p_\alpha} f = {(u_\nu p^\nu)\over \tau_R}(f - f_{\rm leq})
\label{eq:AW_equation}
\end{equation}
Since the metric perturbation is introduced for the sole purpose of obtaining the
stress-energy tensor correlators, it can be considered as an external tensor field in a
flat space defined by $\eta_{\mu\nu} = \hbox{diag}(-1,1,1,1)$.
Therefore, the lowering and raising of the indices are performed with just $\eta_{\mu\nu}$.
Here $\tau_R$ is the relaxation time and
the flow velocity is denoted by $u^\mu$.
The Christoffel symbol $\Gamma^{\alpha}_{\beta\gamma}$ is given by
\begin{equation}
\Gamma^{\alpha}_{\beta\gamma} 
=
{1\over 2}\eta^{\alpha\sigma}
\left(
\partial_\beta h_{\gamma\sigma}
+
\partial_\gamma h_{\beta\sigma}
-
\partial_\sigma h_{\beta\gamma}
\right)
+ O(h^2)
\end{equation}

We use the Maxwell-Boltzmann statistics throughout this study so that
the local equilibrium density $f_{\rm leq}(x,p) = e^{\beta(x) u_\nu(x) p^\nu}$
where $T = 1/\beta$ is the local temperature and $u^\mu$ is the flow velocity.

The local temperature and the flow velocity are 
defined by the Landau matching condition
\begin{equation}
 T^{\mu\nu} u_\nu =  T_{\rm leq}^{\mu\nu} u_\nu = -\varepsilon u^\mu
\end{equation}
with the free-particle equation of state.
Here the kinetic theory definition of the stress-energy tensor
\begin{equation}
T_{\rm a}^{\mu\nu} = 2\int {d^4p\over (2\pi)^4} p^\mu p^\nu\, f_{\rm
a}(x,p)\theta(p^0)(2\pi)\delta(p_\mu p^\mu + m^2)
\end{equation}
is used for any density function labeled a.
Assuming that the metric perturbation is introduced in the global equilibrium background,
we also introduce the equilibrium density
\begin{equation}
f_{\rm eq}(p) = e^{-\beta_0 p^0}
\end{equation}
where $\beta_0 = 1/T_0$ is a global constant.
Up to this point, all density functions should be regarded as functions of $p^\mu$
without enforcing the on-shell condition.

Introducing the deviations 
\begin{equation}
\delta f = f - f_{\rm eq}
\ \hbox{and}\ 
\delta f_{\rm leq} = f_{\rm leq} - f_{\rm eq}
\end{equation}
and regarding all deviations from the global equilibrium
including $\bf u$
to be of $O(h^{\alpha\beta})$,
the kinetic equation can be linearized as
\begin{align}
\mbox{} & p^\mu\partial_\mu \delta f(x,{\bf p}) 
+\beta_0 \Gamma^0_{\beta\gamma}{p^\beta p^\gamma\over E_p} f_{\rm eq}({\bf p})
\nonumber\\
& =
-{E_p\over \tau_R}(\delta f(x,{\bf p}) - \delta f_{\rm leq}(x,{\bf p}))
\end{align}
where the on-shell condition is now used to replace $p^0 \to E_p = \sqrt{p^2 + m^2}$
with $p = |{\bf p}|$.

Using
\begin{equation}
\delta f_{\rm leq}
=
f_{\rm eq}(\beta_0^2 \delta T E_p + \beta_0 {\bf u}\cdot{\bf p})
\end{equation}
and going to the Fourier space, we then obtain
\begin{align}
\lefteqn{
\delta f_{\rm {\bf p}}(\omega, {\bf k})
\left( 1- i\tau_R\omega  + i\tau_R{\bf k}\cdot{\bf v}\right)
}&&
\nonumber\\
& =
E_p\beta_0\left(
(\delta T/T_0) 
+
{\bf u}\cdot{\bf v}
-
\tau_R \Gamma^0_{\beta\gamma}v^\beta v^\gamma
\right)f_{\rm eq}({\bf p})
\end{align}
where
\begin{equation}
v^\mu = p^\mu/E_p = (1, {\bf v})
\end{equation}
The stress-energy tensor deviation can be obtained as follows
\begin{align}
\delta T^{\mu\nu}
& =
\int {d^3p\over (2\pi)^3 E_p} p^\mu p^\nu \delta f
\nonumber\\
& = 
\int {d^3p\over (2\pi)^3 T_0 } f_{\rm eq}({\bf p}) E_p^2 
{
v^\mu v^\nu N
\over
 1- i\tau_R\omega  + i\tau_R{\bf k}\cdot{\bf v}
}
\label{eq:deltaTmunu}
\end{align}
where
\begin{equation}
N
=
v_s^2 \delta T^{00}/h_0
+
v_i \delta T^{0i}/h_0
-
\tau_R \Gamma^0_{\beta\gamma}v^\beta v^\gamma
\end{equation}
To obtain Eq.(\ref{eq:deltaTmunu}), we used the first order relationships
\begin{equation}
\delta T/T_0
=
v_s^2 \delta T^{00}/h_0
\end{equation}
and
\begin{equation}
u^i = \delta T^{0i}/h_0
\end{equation}
where
$v_s$ is the speed of sound and $h_0 =\varepsilon_0 + P_0$ is the enthalpy
density at $T_0$.

Separating out the angle integrals, we can see that the right-hand-side 
of Eq.(\ref{eq:deltaTmunu}) can be built out
of the following integrals
\begin{align}
\Omega^{i_1\cdots i_l}(\omega, k_z)
=
{1\over 2\pi^2 T}\int_0^\infty dp\, e^{-E_p/T_0}\, E_p^2 p^2\, v^l N^{i_1\cdots i_l} 
\end{align}
where $v = p/E_p$, and
\begin{equation}
N^{i_1\cdots i_l}(\omega, k_z, v)
=
\int {d\Omega_n\over (4\pi)}
{n^{i_1} n^{i_2}\cdots n^{i_l}
\over 
 1- i\tau_R\omega  + i\tau_R{\bf k}\cdot{\bf v}
}
\end{equation}
with ${\bf n} = {\bf v}/v$, $n^0 = 1$, and $0 \le l \le 4$.
Here $i_n$'s are spatial indices.
In the massless limit where $m\to 0$, $v\to 1$, the angle integral and the energy integral
separate to give
\begin{align}
\Omega^{i_1\cdots i_l}
& =
3 h_0 N^{i_1\cdots i_n}
\end{align}
But with $m\ne 0$, the two integrals are tangled.

To simplify further, we assume that $h^{\mu\nu} = h^{\mu\nu}(t, z)$ and hence ${\bf k} =
k_z{\bf e}_z$.
With that choice,
$N^{i_1\cdots i_l}$ with odd number of 1's or 2's in the indices
vanish. For instance, $N^{1222} = N^{1123} = 0$, but $N^{1122} \ne 0$.
Since the denominator of $N^{n_1\cdots n_i}$ does not involve the azimuthal angle
$\phi$,
any appearance of $(11)$ or $(22)$ can be replaced by $\pi$.
For instance, $N^{11} = N^{22} = \pi N^{0}$, and
$N^{113} = N^{223} = \pi N^{3}$. Any appearance of $(1122)$ or its permutation can be replaced
by $\pi/4$. Hence $N^{1122} = N^{1212} = (\pi/4)N^0$. Any appearance of $(1111)$ or
$(2222)$ can be replaced by $3\pi/4$ so that $N^{1111} = N^{2222} =
(3\pi/4)N^0$.

The stress-energy tensor deviation with the non-zero mass are given by 
\begin{align}
\delta T^{00} 
&= 
-\frac{h_0 (S^{00} (h_0-\Omega^{33})+\Omega^{3} S^{03})}
{-h_0 (\Omega^{33}+\Omega^{0} v_s^2)+v_s^2 \left(\Omega^{0} \Omega^{33}-(\Omega^{3})^2\right)+h_0^2}
\\
\delta T^{03} 
&= 
-\frac{h_0 (S^{03} (h_0-\Omega^{0} v_s^2)+\Omega^{3} v_s^2 S^{00})}
{-h_0 (\Omega^{33}+\Omega^{0} v_s^2)+v_s^2 \left(\Omega^{0} \Omega^{33}-(\Omega^{3})^2\right)+h_0^2}
\\
\delta T^{01}
&=
{S^{01}\over \Omega^{11}/h_0 - 1}
\\
\delta T^{02}
&=
{S^{02}\over \Omega^{22}/h_0 - 1}
\end{align}
and
\begin{align}
\delta T^{ij}
&=
{v_s^2 \delta T^{00}\over h_0} \Omega^{ij}
+
{\delta T^{0k}\over h_0}\Omega^{ijk}
-S^{ij}
\end{align}
where $i$ and $j$ are spatial indices.
The source terms are
\begin{align}
S^{00}
&=
\tau_R \Gamma_{00}^0\Omega^0 + 2\tau_R\Gamma_{0i}^0\Omega^i
+ \tau_R\Gamma^0_{ij}\Omega^{ij}
\\
S^{0i}
&=
\tau_R \Gamma_{00}^0\Omega^i + 2\tau_R\Gamma_{0j}^0\Omega^{ij}
+ \tau_R\Gamma^0_{jk}\Omega^{ijk}
\\
S^{ij}
&=
\tau_R \Gamma_{00}^0\Omega^{ij} + 2\tau_R\Gamma_{0k}^0\Omega^{ijk}
+ \tau_R\Gamma^0_{kl}\Omega^{ijkl}
\end{align}
The correlation function are then obtained using 
\begin{align}
\delta T^{\alpha\beta} = -{1\over 2}G^{\alpha\beta,\mu\nu} h_{\mu\nu} + O(h^2)
\end{align}

All stress-energy tensor correlation functions can be expressed in terms of five
independent response functions defined in Ref.\cite{Jeon:2025kem}.
The following response functions are all functions of
$(\omega, k_z)$.
\begin{align}
G_L
& = {\omega^4\over k_z^4}G^{00,00} - {\omega^2\over k_z^2}h_0
\nonumber\\
& =
-\frac{h_{0} \omega^2}{k_z^2}
\nonumber\\
+\, &
\frac{i h_0 \omega ^4 \tau_R \left( 
h_0 \omega  \Omega^{0}
+\omega (\Omega^{3})^2
-\omega  \Omega^{0} \Omega^{33}
-2 h_0 k_z \Omega^{3}
\right)}{k_z^4 \left(-h_0 (\Omega^{33}+\Omega^{0} v_s^2)+v_s^2
   \left(\Omega^{0} \Omega^{33}-(\Omega^{3})^2\right)+h_0^2\right)}
\\
G_{LT}
& =
{\omega^2\over k_z^2}(G^{11,00} + G^{00,11})/2
\nonumber\\
& =
\frac{-i h_0 \omega^2 \tau_R (h_0 \Omega^{113}+\Omega^{3} v_s^2
   \Omega^{11}-\Omega^{0} v_s^2 \Omega^{113})}{k_z \left(-h_0 \Omega^{33}+\Omega^{0}
   v_s^2 \Omega^{33}+h_0^2-h_0 \Omega^{0} v_s^2-(\Omega^{3})^2 v_s^2\right)}
\\
G_T
& =
(G^{11,22} + G^{22,22})/2
\nonumber\\
& =
{N_T
 \over
 2 \left(-h_0 (\Omega^{33}+\Omega^0 v_s^2)+v_s^2 
 \left(\Omega^{0} \Omega^{33}-(\Omega^{3})^2\right)+h_0^2\right)
 }
\\
N_T
& =
-i \omega  \tau_R \Big(h_0^2 (\Omega^{1111}+\Omega^{1122})
+h_0 \big(2 v_s^2 (\Omega^{11})^2
\nonumber\\ &\quad
-\Omega^{0} v_s^2 (\Omega^{1111}+\Omega^{1122})+2 (\Omega^{113})^2
\nonumber\\ &\quad
-\Omega^{33} (\Omega^{1111}+\Omega^{1122})\big)
\nonumber\\ &\quad
+v_s^2 \big(-\left( (\Omega^{3})^2 (\Omega^{1111}+\Omega^{1122})\right)+4 \Omega^{3} \Omega^{11} \Omega^{113}
\nonumber\\ &\quad
-2 (\Omega^{11})^2 \Omega^{33}
\nonumber\\ &\quad
+\Omega^{0} \left(\Omega^{33} (\Omega^{1111}+\Omega^{1122})-2 (\Omega^{113})^2\right)\big)\Big)
\\
G_1 
& = G^{2323}
=
-i \omega  \tau_R \left(\frac{(\Omega^{113})^2}{h_0-\Omega^{11}}+\Omega^{1133}\right)
\\
G_2 
& = G^{1212}
=
-i \omega  \tau_R \Omega^{1122}
\end{align}

Similar expressions for the
correlation functions in the massless theory can be found in Refs.
\cite{Romatschke:2015gic, Bajec:2024jez}.
For the massive theory,
some are listed in Refs.\cite{
Hataei:2025mqf,
Bajec:2025dqm}.
For completeness, we briefly illustrate how to evaluate $N_n^{i_1\cdots i_n}$
in Appendix \ref{app:evalN} in the small $m/T$ limit.

\section{New Kubo formulas}

The kinetic theory
response functions obtained in the previous section satisfy the small $\omega$ limit
requirements \cite{Jeon:2025kem}
\begin{align}
G_L(\omega,k^2) & = -h_0{\omega^2\over k^2} - {\omega^4\over k^4}g_L(k^2) + O(\omega^5)
\end{align}
\begin{align}
G_{LT}(\omega,k^2) & = -h_0{\omega^2\over k^2} - \omega^2 g_{LT}(k^2) + O(\omega^3)
\end{align}
\begin{align}
G_{1}(\omega,k^2) & = -h_0{\omega^2\over k^2} - \omega^2 g_{1}(k^2) + O(\omega^3)
\end{align}
They also satisfy the small $k$ limit requirements
\begin{align}
2G_1 - G_L + G_{LT} & = O(k^2)
\\
2G_T - G_{LT} + G_{L} & = O(k^2)
\\
G_2 - 4G_1 + G_T + G_L - 2G_{LT} & = O(k^4)
\end{align}
Although they cannot get the thermodynamic coefficients $\kappa$ and $\kappa^*$,
fulfilling both the small $\omega$ and the small $k$ behaviours is a solid indication that these
kinetic theory response functions have the required analytic structures, and hence can be
used to evaluate the new Kubo formulas.

In Ref.\cite{Jeon:2025kem}, eight new Kubo formulas were introduced. 
For the following seven, the $\omega\to 0$ limit is taken first
\begin{align}
\eta & = {i\over 12}\lim_{k\to 0}\lim_{\omega\to 0}\Delta_{ij,lm}\partial_\omega
G^{ij,lm}(\omega, {\bf k})
\label{eq:dGDPDP}
\\
\eta/h_0 
& =
-i\lim_{k_z\to 0}\lim_{\omega\to 0} {h_0\over k_z^2\partial_\omega G_R^{01,01}(\omega, k_z)}
\label{eq:dG0101_1}
\\
\eta/h_0
& =
\left(
\lim_{k_z\to 0}\lim_{\omega\to 0}
{2h_0\over k_z^4\partial_\omega^2 G_R^{01,01}(\omega, k_z)}
\right)^{1/2}
\label{eq:dG0101_2}
\\
\eta
& =
{i\over 4}
\lim_{k_z\to 0}\lim_{\omega\to 0}\partial_\omega G^{11,11}(\omega, k_z)
\label{eq:dG1111}
\\
\left(\zeta + 4\eta/3\right)/v_s^4
& =
i 
\lim_{k\to 0}\lim_{\omega\to 0}\partial_\omega G^{\epsilon\epsilon}(\omega, k)
\label{eq:dGepseps}
\\
\eta/v_s^2 & =
{3i\over 4} 
\lim_{k\to 0}\lim_{\omega\to 0}\partial_\omega G^{\epsilon P}(\omega, k)
\label{eq:dGepsP}
\\
\eta & =
{3i\over 4} 
\lim_{k\to 0}\lim_{\omega\to 0}\partial_\omega G^{P P}(\omega, k)
\label{eq:dGPP}
\end{align}
where we defined the energy density operator $\hat\epsilon = \hat T^{00}$ and
the pressure operator $\hat P = \hat T^{ii}/3$ in the global rest frame.
Here
\begin{align}
\Delta^{ij,lm} & = {1\over 2}
\left(
\Delta^{il}\Delta^{jm} + \Delta^{im}\Delta^{jl} - (2/3)\Delta^{ij}\Delta^{lm}
\right)
\end{align}
is the spin-2 projector in terms of the local 3-metric $\Delta^{ij} = g^{ij} + u^i u^j$
and the flow velocity $u^\mu = (1,0,0,0)$.
The last three can be combined to yield
\begin{align}
\zeta & = i\lim_{k\to 0}\lim_{\omega\to 0}\partial_\omega G^{\Delta P\Delta P}
\end{align}
where $\Delta\hat P = \hat{P} - v_s^2\hat\epsilon$.
For this correlation function, taking the $k\to 0$ limit first also gives $\zeta$.

If we reverse the order of the limits, the right hand side (RHS) Eq.(\ref{eq:dGDPDP}) yields $5\eta/3$,  
the RHSs of Eqs.(\ref{eq:dG0101_1}) and (\ref{eq:dG0101_2}) do not exist, the RHS of
Eq.(\ref{eq:dG1111}) yields our eighth Kubo formula,
\begin{align}
\left(\zeta + 4\eta/3\right)
& =
i
\lim_{\omega\to 0}\lim_{k_z\to 0}\partial_\omega G^{11,11}(\omega, k_z)
\end{align}
the RHS of Eq.(\ref{eq:dGepseps}) and Eq.(\ref{eq:dGepsP}) vanish in this limit, and
the RHS of Eq.(\ref{eq:dGPP}) yields $(3/4)\zeta$.

To see which one of the five independent correlation functions contribute to each of these
formula, consider first the transverse-transverse
correlation functions that appears in the standard
the shear viscosity formula:
\begin{align}
G^{TT} & = \Delta_{ij,lm}G_R^{ij,lm}
\end{align}
From the decomposition of $G_R^{ij,lm}$ illustrated in Ref.\cite{Jeon:2025kem}, one can see
\begin{align}
G^{TT} = 8(G_1 + G_2) + (4/3)(G_L + G_T - 2G_{LT})
\end{align}
where we have omitted the contact term contribution (we will do so in the remainder of
this section)
as it does not contribute to the Kubo formulas.
If we take the $k\to 0$ limit first, we obtain
\begin{align}
\partial_\omega G^{TT}(\omega, 0)
& =
20 \partial_\omega G_1(\omega,0) 
\label{eq:GTT_lim_k}
\end{align}
where we used the small $k$ relationships between the correlation functions given in
the last section.
On the other hand,
\begin{align}
\partial_\omega G^{TT}(0,k)
& =
8\partial_\omega  G_2(0,k)
+
(4/3)\partial_\omega G_T(0,k) 
\end{align}
For Eq.(\ref{eq:dG0101_1}) and Eq.(\ref{eq:dG0101_2}), we have
\begin{align}
G^{01,01}_R(\omega, k_z) = {k_z^2\over \omega^2}G_1(\omega, k_z)
\end{align}
For Eq.(\ref{eq:dG1111}), we can show
\begin{align}
G^{11,11}_R(\omega, k_z) = G_2(\omega, k_z) + G_T(\omega, k_z)
\end{align}
Both terms contribute differently in the two different orderings of the limits.

For the longitudinal correlators, we have
\begin{align}
G^{\epsilon\epsilon} & = {k^4\over \omega^4}G_L + {k^2\over \omega^2}h_0
\label{eq:Gepseps}
\\
G^{\epsilon P} & =
{k^2\over 3\omega^2}\left(G_L + 2G_{LT}\right)
\\
G^{PP}
& =
(1/9)(G_L + 4G_{LT} + 4G_T)
\end{align}
Taking the $\omega$ derivative, we then obtain
\begin{align}
\partial_\omega G^{\epsilon\epsilon}(\omega, 0) & = & 0 
\\
\partial_\omega G^{\epsilon P}(\omega, 0) & = & 0
\\
\partial_\omega G^{PP}(\omega, 0) & = &
\partial_\omega\left( G_L - (4/3) G_1\right)_{k\to 0}
\end{align}
again taking into the small $k$ relationships between the correlation functions.
On the other hand,
\begin{align}
\partial_\omega G^{\epsilon\epsilon}(0,k)
& =
\partial_\omega
\left(
{k^4\over \omega^2}G_L + {k^2\over \omega^2}h_0
\right)_{\omega\to 0}
\\
\partial_\omega G^{\epsilon P}(0,k) & =
\partial_\omega \left( {2k^2\over 3\omega^2}G_{LT} \right)_{\omega \to 0}
\\
\partial_\omega G^{PP}(0,k)
& =
(4/9)\partial_\omega G_T(0,k)
\end{align}
Therefore, to use the new Kubo formulas, one need to know all five response functions with
enough accuracy.

\section{Minimum requirements}

From the analysis in Ref.\cite{Jeon:2025kem}, one can determine the minimum form of the five correlation
functions that satisfies all Kubo formulas in the last section.
These are\footnote{In Ref.\cite{Jeon:2025kem}, there was a typographical error in the expressions of $G_L,
G_{LT}$ and $G_T$. The sign of $R(\omega,k)$ should have been $+$.}
\begin{align}
G_1 & =
-{h_0\omega^2\over k^2 - i\omega/D_R}
\\
G_2 & =
-ih_0\omega D_R
\\
G_L & =
{\omega^2 h_0 (Z_R - i\omega Z_I)\over 
\omega^2 - (Z_R - i\omega Z_I)k^2 - i\omega Z_R R_R
}
\\
G_{LT}
& =
{\omega^2(Z_R - i\omega Z_I)(h_0 + i\omega F_{LTR})
\over
\omega^2 -(Z_R - i\omega Z_I) k^2 -i\omega^3 Z_R R_R
}
\end{align}
\begin{align}
G_{T}
& =
{
(Z_R - i\omega Z_I)\left( \omega^2( h_0 + i\omega F_{T1R}) + i\omega k^2 F_{T2R} \right)
\over
\omega^2 -(Z_R - i\omega Z_I) k^2 -i\omega^3 Z_R R_R
}
\end{align}
where the coefficients are given by
\begin{align}
D_R & = \eta/h_0
\\
Z_R & = v_s^2
\\
Z_I & = (\zeta + 4\eta/3)/h_0 + Z_R^2 R_R
\\
F_{LTR} & = {2\eta\over v_s^2}
\\
F_{T1R} & = {\eta\over v_s^2} 
\\
F_{T2R} & = 3\eta
\end{align}
The kinetic theory response functions indeed satisfy these requirements.
In Ref.\cite{Jeon:2025kem}, it was shown that $R_R$ is the sum of all relaxation times.
Therefore, the sound attenuation constant $Z_I$ contains both the contribution from the
viscosities and the relaxation times.
It was also shown that the value of $R_R$ is ambiguous since its value depended on 
how many relaxing modes are added to the hydrodynamics.
It is then interesting to see that the kinetic theory correlators has $v_s^2 R_R = 6\tau_R$. 
With the same relaxation time approximation, the second order hydrodynamics derived from
the kinetic theory would result in $v_s^2 R_R = 2\tau_R$ (with $m\ne 0$), and
the third order hydrodynamics with $m = 0$ would result 
in $v_s^2 R_R = 3 \tau_R$.
Since the massive second order hydrodynamics has 2 relaxing modes and the massless third
order hydrodynamics has 3 relaxing modes, one may speculate that the quantity
$n_R = v_s^2/(\tau_R R_R)$
represents the total number of relaxing modes.
However, at this point, we do not know whether there are 6 (and only 6) relaxing modes
contained in these response functions.

\section{Verifying the new Kubo formulas}

First, consider the massless case.
Since $m^2 = 0$, ${\bf v} = {\bf p}/p$, and $I_n^{i_1\cdots i_n}$
does not depend on $p$. 
Then all integrals can be carried out analytically.
To verify the new Kubo formulas 
from Ref.\cite{Jeon:2025kem} for the massless case,
we need to know all five independent response functions.
Among them, three ($G_1, G_2, G_L$ in \cite{Jeon:2025kem}) were calculated in 
Refs.\cite{Romatschke:2015gic}
the rest ($G_{LT}$ and $G_T+G_2$) were calculated in Ref.\cite{Bajec:2024jez}.
Using their results, one can then verify all new Kubo formulas 
with the help of symbolic manipulation programs such as {\sc Mathematica}.
One then recovers
the vanishing bulk viscosity $\zeta = 0$, the massless theory speed of sound $v_s^2
=1/3$, and the specific shear viscosity $\eta/h_0 = \tau_R/5$. 

For $m^2 \ne 0$, 
some of the correlation functions
are listed in Refs.\cite{Hataei:2025mqf,Bajec:2025dqm}.
We have listed our own solutions in section \ref{sec:non_zero_m_kinetic} as well.
Again, with the help of symbolic manipulation programs, one can verify that the new Kubo
formulas in the small $m/T$ limit all result in
\begin{align}
\eta/h_0 & = 
\tau_R\left(
{1\over 5} - {1\over 60}\xi^2 + {1\over 96}\xi^4
\right)
+ O(\xi^5)
\\
\zeta/h_0 & =
{5\over 432}\tau_R \xi^4 + O(\xi^5)
\end{align} 
where $\xi = m/T$.
These agree with with the results from Ref.\cite{Bajec:2025dqm}, and 
also numerically agree with the results from Ref.\cite{Denicol:2014vaa}. 
Comparing with the results in Refs.\cite{Hataei:2025mqf}, the shear viscosity result agrees
up to and including $O(\xi^2)$, but the bulk viscosity result does not agree. 
The enthalpy density and the speed of sound are known to be
\begin{align}
h_0 
& =
{T^4\over 16\pi^2}
\left( 64 - 8 \xi^2 + \xi^4 \right)
+ O(\xi^6\log\xi)
\nonumber\\
v_s^2 & = 
\frac{1}{3} -\frac{\xi^2}{36} +\frac{5 \xi^4}{864} + O(\xi^6\log\xi)
\end{align}

\section{Summary \& Discussions}

In this work, we have verified the new Kubo formulas that were first demonstrated in
Ref.\cite{Jeon:2025kem} using the stress-energy tensor correlation functions from kinetic
theory. There were two main purposes for doing so. The first is to show the veracity and
consistency of the new Kubo formulas with the reversed order of the limits
within a simplest non-trivial setting.
The second is to show that the analytic structures
of the correlation functions derived in Ref.\cite{Jeon:2025kem} are indeed realized in a
kinetic theory setting.
Since the ladder diagram resummation can be re-organized to recover linearized kinetic
theory,
this is an important intermediate step and consistency check
towards a full field theory calculation.
Remarkably, all eight Kubo formulas give the consistent values for 
$\eta/h_0$ and $\zeta/h_0$,
and we believe that this work demonstrates the potential utility of the new Kubo formulas.

In most of the new Kubo formulas, the $\omega\to 0$ limit is taken first, and then the
$k\to 0$ limit is taken. On the surface, this may look like the static limit of 
correlation functions. However, due to the presence of frequency derivatives, this is not
really the case. Instead, the new Kubo formulas come from structural and analytic
consistency between
stress-energy tensor correlation functions. We stress that these consistency conditions
are completely general. However, this does imply
that direct lattice implementation would be
difficult to achieve.

One natural extension of this work would be to include the space-time dependent
quasiparticle mass which would add a term proportional to
${\partial m^2\over \partial x_\mu}{\partial\over \partial p^\mu} f$ to
Eq.(\ref{eq:AW_equation}). This is a non-trivial problem
\cite{Biro:1990vj,
Gorenstein:1995vm,
Peshier:1995ty,
Jeon:1994if,
Hong:2010at,
Chakraborty:2010fr,
Czajka:2017bod,
Czajka:2017wdo,
Rocha:2022fqz}
due to the fact that so far no unique way has been found 
to define a conserved stress-energy tensor within the relaxation time approximation 
that is also compatible with the Landau condition (or any frame choice). 
Another natural extension would be to work this out in a simple field theory such as the
scalar theory via the effective Boltzmann equation.
Investigation on these topics is underway.

\acknowledgments

S.J.~acknowledge the support of the Natural Sciences and Engineering Research Council of Canada (NSERC),
[SAPIN-2024-00026]. S.J.~also thanks Jingxuan (Amelia) Feng who helped checking the results for
the massless theory.
A.C.~is supported in part by the National
Science Centre (Poland) under the research Grant No.~2021/43/D/ST2/01154 (SONATA 17). 
J.H.~is supported by the National Research Foundation of Korea (NRF)
grant funded by the Korean government (MSIT) (No.~RS-2024-00342514).

\bibliography{kubo_refs}

\appendix

\section{Evaluation of $N^{i_1\cdots i_l}$ in the small $m/T$ limit}
\label{app:evalN}

\begin{align}
N^{i_1\cdots i_l}
& =
\int {d\Omega_v \over 4\pi}
{n^{i_1} \cdots n^{i_m}\over 1 - i\tau_R \omega + i\tau_R k_z v\nu}
\end{align}
where ${\bf n} = {{\bf v}\over |{\bf v}|} = (\sin\theta\cos\phi, \sin\theta\sin\phi, \cos\theta)$
is the unit vector in the direction of ${\bf v}$, and $\nu = \cos\theta$.
We can use
\begin{align}
v 
=
{p\over E_p}
=
\sqrt{1 - m^2/E_p^2}
\end{align}
to expand the denominator in powers of $m/E_p$ resulting in
\begin{align}
\lefteqn{
{1\over 1 - i\tau_R \omega + i\tau_R k_z v}
}&
\nonumber\\
& =
{1\over i\tau_R k_z}
\left(
{1\over A + \nu}
+ {m^2\nu\over 2E_p^2(A + \nu)^2}
+ {m^4(A\nu + 3\nu^2)\over 8E_p^4(A + \nu)^3}
\right)
\nonumber\\ &\quad
+ O(m^6/E_p^6)
\end{align}
with $A = (1-i\tau_R\omega)/(ik_z\tau_R)$.
This expansion
is adequate to get the leading order shear viscosity and the bulk viscosity
since it is known that $\eta = O(\tau_R T^4)$ and $\zeta = O(\tau_R m^4)$.
One can then define
\begin{align}
I^{i_1\cdots i_l} 
&=
{1\over i\tau_Rk_z}
\int {d\Omega\over 4\pi}{n^{i_1}\cdots n^{i_l}\over A + \nu}
\end{align}
\begin{align}
J^{i_1\cdots i_l} 
&=
{1\over i\tau_Rk_z}
\int {d\Omega\over 4\pi}{n^{i_1}\cdots n^{i_l}\nu\over 2(A + \nu)^2}
\end{align}
\begin{align}
K^{i_1\cdots i_l} 
&=
{1\over i\tau_Rk_z}
\int {d\Omega\over 4\pi}{n^{i_1}\cdots n^{i_l}(A\nu+3\nu^2)\over 8(A + \nu)^3}
\end{align}
to get
\begin{align}
\Omega^{i_1\cdots i_l}
\approx
P_{l}I^{i_1\cdots i_l}
+
m^2 Q_{l} J^{i_1\cdots i_l}
+
m^4 R_{l} K^{i_1\cdots i_l}
\end{align}
Here we defined
\begin{align}
P_l
& =
{1\over 2\pi^2 T}
\int_0^\infty dp\, E_p^2 p^2\, (p/E_p)^l e^{-E_p/T}
\end{align}
\begin{align}
Q_l
& =
{1\over 2\pi^2 T}
\int_0^\infty dp\, p^2\, (p/E_p)^l e^{-E_p/T}
\end{align}
and
\begin{align}
R_l
& =
{1\over 2\pi^2 T}
\int_0^\infty dp\, (p/E_p)^{l+2} e^{-E_p/T}
\end{align}
When $l$ is 1 or 3, these integrals can be evaluated explicitly using $dp(p/E_p) = dE_p$.
For $l = 0, 2, 4$, 
these integrals can be evaluated using the modified Bessel function
\begin{align}
K_n(\xi) = \left( \sqrt{\pi}(\xi/2)^n\over \Gamma(n+1/2)\right)\int_0^\infty dy\,
\sinh^{2n}(y)\, e^{-\xi\cosh(y)}
\label{eq:BesselK_n}
\end{align}
and the Bickley-Naylor $\hbox{Ki}$ function
\begin{align}
\hbox{Ki}_n(\xi) = \int_0^\infty dy\, {e^{-\xi\cosh y}\over \cosh^n(y)}
\end{align}
with $\xi = m/T$ and using $p = m\sinh(y)$.

To get the leading order shear viscosity and the bulk viscosity, one needs to know $P_l$
up to and including $O(m^4)$, $Q_l$ up to and including $O(m^2)$, and 
\begin{align}
R_l \approx {1\over 2\pi^2}
\end{align}

\end{document}